# Cost-Benefit Analysis using Modular Dynamic Fault Tree Analysis and Monte Carlo Simulations for Condition-based Maintenance of Unmanned Systems

Joseph M. Southgate[1], Katrina Groth[2], Peter Sandborn[3], Shapour Azarm[4]

Center for Risk and Reliability, Department of Mechanical Engineering
University of Maryland
College Park, Maryland, 20742 USA
[1]Corresponding author: jsouthgate4@gmail.com; {kgroth[2], sandborn[3], azarm[4]}@umd.edu

## ABSTRACT

Recent developments in condition-based maintenance (CBM) have helped make it a promising approach to maintenance cost avoidance in engineering systems. By performing maintenance based on conditions of the component with regards to failure or time, there is potential to avoid the large costs of system shutdown and maintenance delays. However, CBM requires a large investment cost compared to other available maintenance strategies. The investment cost is required for research, development, and implementation. Despite the potential to avoid significant maintenance costs, the large investment cost of CBM makes decision makers hesitant to implement.

This study is the first in the literature that attempts to address the problem of conducting a cost-benefit analysis (CBA) for implementing CBM concepts for unmanned systems. This paper proposes a method for conducting a CBA to determine the return on investment (ROI) of potential CBM strategies. The CBA seeks to compare different CBM strategies based on the differences in the various maintenance requirements associated with maintaining a multi-component, unmanned system. The proposed method uses modular dynamic fault tree analysis (MDFTA) with Monte Carlo simulations (MCS) to assess the various maintenance requirements. The proposed method is demonstrated on an unmanned surface vessel (USV) example taken from the literature that consists of 5 subsystems and 71 components. Following this USV example, it is found that selecting different combinations of components for a CBM strategy can have a significant impact on maintenance requirements and ROI by impacting cost avoidances and investment costs.

## 1. INTRODUCTION

Condition-based maintenance (CBM) is a strategy for improving maintenance practices and ultimately reducing maintenance costs in engineering systems. CBM concepts use prognostics and health management (PHM) systems, structural health monitoring (SHM) systems or condition monitoring systems (CMS) to conduct maintenance based on the component's conditions prior to failure. Unmanned systems, with no personnel onboard or available during operations, can utilize CMS to trigger maintenance for the periods when personnel are available to conduct maintenance. Many asset-intensive organizations already ojincorporate some degree of CBM in the maintenance schemes of their engineering systems, including manufacturing (Prajapati, Bechtel, and Ganesan, 2012; Rastegari & Bengtsson, 2014; Berdinyazov, Camci, Baskan, Sevkli, and Eldemir, 2011; Singh & Verma, 2020), automotive and aerospace (Prajapati et al. 2012; Berdinyazov et al. 2011), military (Prajapati et al. 2012), and power plant industry (Berdinyazov et al. 2011).

Despite many possible benefits from CBM, organizations remain hesitant in adopting it because of high initial investment and development costs (Prajapati et al. 2012; Rastegari & Bengtsson, 2014; Berdinyazov et al. 2011). Berdinyazov et al. (2011) state that: "An analysis tool that will evaluate the possible value of CBM by comparing corrective maintenance (CM) and preventative maintenance (PM) considering its investment, setup and management costs…is now a great need for industry." Yoon, Youn, Yoo, Kim, and Kim (2019) state that: "...maintenance cost analysis based on fault diagnosis has not been well investigated." Our paper is motivated by the need to develop an analysis tool to enable decision makers to have a better understanding of how CBM will impact the lifetime costs associated with maintaining their system and thus decide whether CBM is a cost-effective investment, especially when considering unique maintenance challenges of unmanned systems.

Deciding if CBM is a good investment can be accomplished through several CBA techniques. Examples of CBA techniques used in the literature include cost avoidance (CA), ROI (Sandborn & Lucyshyn, 2023), life cycle costs (LCC) (Sandborn & Lucyshyn, 2023; Torti, Venanzi, Laflamme, and Ubertini, 2022), economic analysis (Berdinyazov et al. 2011), cost-effectiveness analysis (CEA) (De Carlo & Arleo, 2013; Rastegari & Bengtsson, 2015). When considering CBM, such techniques involve performing some form of comparison between the maintenance costs of adding and using CBM and the maintenance costs without CBM. The comparison typically involves comparing the number of failures and CM activities to the number of PM actions of one maintenance strategy against the number of CM and PM activities of another. CBM seeks to identify failures early and conduct PM before the failure occurs (Prajapati et al. 2012; Teixeira, Lopes, and Braga, 2020). However, regardless of the cost analysis technique used, CBA studies do not consider several challenges facing unmanned systems, such as lack of personnel during operation periods to conduct maintenance (Komianos, 2018; Chang, Kontovas, Yu, and Yang, 2021; Dreyer & Oltedal, 2019). These periods where maintenance is unavailable raise the system's dependency on non-critical components and the probability of system failure (Chang et al. 2021; Dreyer & Oltedal, 2019). Whether or not a components failure results in system failure has significant implications for costs and must be considered in a CBA.

In this paper, an approach for CBA of a CBM strategy for an unmanned system is proposed. The approach consists of a MDFTA using MCS (Sandborn, & Lucyshyn, 2023; Gascard & Simeu-Abazi, 2018; Rao, Gopika, Rao, Kushwaha, Verma, and Srividya, 2009; Gulati & Dugan, 1997; Aslansefat, Kabir, Gheraibia, and Papadopoulos, 2020) to approximate the system's maintenance requirements under different maintenance strategies over a predetermined operating period. Using this approach, the system's maintenance requirements are established based on several factors which are evaluated by the MDFTA. The maintenance requirements of the system and the investment costs of the CMS are then used to determine the CA and ROI of the new maintenance strategy and compared to the system's current maintenance strategy. These comparisons allow for decision makers to identify the better CBM strategies among potential options. There are varying levels of autonomy in the literature, but this study will focus on fully autonomous or unmanned systems (Komianos, 2018). The proposed approach has been demonstrated with an application to a USV case study (Gao, Guo, Zhong, Liang, Wang, and Yi, 2021).

The main contributions of this paper in view of previous related works are as follows:

- The proposed approach considers unique maintenance challenges facing unmanned systems which may affect cost. The challenges considered in this study are lack of maintenance personnel while operating, high costs of system failure, and no production value (Komianos, 2018; Chang et al. 2021; Dreyer & Oltedal, 2019; Yang, Vatn, and Utne, 2023). These challenges must be considered when determining the impact of a maintenance strategy on maintenance costs which are not considered in the current studies on CBA of CBM. Specifically, this paper provides a more tailored approach to assess the maintenance needs of this class of engineering systems, i.e., unmanned systems.

- The proposed approach considers the system's degraded operations and how that impacts maintenance costs. Multi-component systems with a mixture of critical and non-critical components can have some component failures which may not result in system failure and the system continues to operate in a third state, the degraded state (De Jonge & Scarf, 2020). Unmanned systems could experience long periods of degraded operations and thus that time needs to be considered, as done in this study, in a cost analysis (Chang et al. 2021; Dreyer & Oltedal, 2019).

- Finally, the proposed approach for CBA of CBM considers the entire system, its individual components, and the impact its component level maintenance can have on other components in the system. Poppe, Boute, & Lambrecht (2018) discuss system-wide consideration by stating how single component CBM studies are useful, but most practical applications will involve multi-component systems. Poppe et al. (2018) further elaborate on how bundling CBM strategies has the potential to reap additional benefits. This study will show how adding CBM to a single component will propagate through the system and alter maintenance requirements of other components in the system.

The remainder of this paper is organized as follows. Section 2 provides a more detailed background and literature review material including types of unmanned systems and their maintenance challenges, type of maintenance and their associated costs, CBM investment costs, CBA models, and MDFTA. Section 3 details the proposed methodology including problem aim, assumptions, and details of the approach used. Section 4 demonstrates the methodology on a USV example from the literature. Section 5 provides discussion and insights into the results of Section 4. Section 6 is the conclusion and highlights areas for improvement in future works.

## 2. BACKGROUND AND ADDITIONAL LITERATURE SURVEY

This section will provide background information on topics related to this study. The provided information will aid in better understanding of the approach and its implications.

### 2.1. Unmanned Systems and Maintenance Challenges

Zhang, Zhang, Liang, Li, Wang, Li, Zhu, & Wu (2017) describe unmanned systems as: "…systems that are man-made and capable of carrying out operations or management by means of advanced technologies without human intervention." Unmanned systems include but are not limited to unmanned vehicles (UV) (Zhang et al. 2017), unmanned ariel vehicles (UAV) (Zhang et al. 2017), drones (Zhang et al. 2017), USV (Gao et al. 2021), MASS (Chang et al. 2021; Dreyer & Oltedal, 2019; Yang et al. 2023), autonomous marine systems (AMS) (Yang et al. 2023), autonomous underwater vehicles (AUV) (Yang et al. 2023), & robotics (Zhang et al. 2017). Using the description of unmanned systems by Zhang et al. (2017), the system operates "without human intervention." However, when the system is not operating, human intervention is required.

Regardless of the title or area of operation of the unmanned system, reliability during operations is essential. During operations, any component failures cannot be repaired until the mission is completed and the system is available to maintenance personnel. Therefore, appropriate maintenance plans for components in the system are required to maintain acceptable levels of reliability. Yang et al. (2023) highlight some difficulties with maintaining unmanned systems. Specifically, Yang et al. (2023) identify high consequences of system failure and limited maintenance opportunities. Limited maintenance opportunities for USVs, MASS, AMS, and AUVs could refer to times when the vehicles are in-port and maintenance personnel are present. With limited opportunities, when maintenance of the system is available, any failed components must be corrected in addition to rigorous PM actions which will serve to ensure the system is prepared for the next operation. The final consideration for unmanned systems in this study is that the system completes repeat missions and does not have an hourly production value. This is unique as most CBA study include the impact of lengthy maintenance actions on lost production value.

## 2.2. Types of Maintenance and Associated Costs

Proper maintenance is necessary to maintain high reliability and safe operations and makes up a significant contribution to the organization's expenses (Zhao, Xu, Liang, Zhang, and Song, 2019). Maintenance actions are classified in one of two types: corrective (CM) and preventative (PM) (Rausand & Hoyland, 2003). CM or breakdown maintenance is performed after a component or system failure and seeks to bring back the failed unit to an operational state again. PM can be further subdivided into age-based, clock-based, and condition-based. Age-based and clock-based maintenance, also known as time-based maintenance (TBM), are two of three subcategories of PM. This study will focus on CM and PM from CBM. Maintenance costs over a system's life can most simply be described as cost per maintenance activity times the number of activities. This simple calculation can provide the total costs of maintaining the system over a desired period. However, the cost per maintenance activity can be broken down into many parts. The literature provides many descriptions of how the cost per maintenance activity can be broken up.

Rastegari and Bengtsson (2015) categorize maintenance costs as either direct or indirect sources. Reducing system damage and expenses from maintenance is a direct cost saving while avoiding losses in production is an example of indirect cost savings. This study will only consider the direct costs for component maintenance. The direct cost considers the cost to cover parts and labor to conduct the required repairs. Indirect costs such as lost production are not considered because this study is focused on unmanned system completing missions and there is no set value for completing the mission or lost costs for lengthy repairs. Studies by: Berdinyazov et al. (2011), De Carlo and Arleo (2013), Rastegari and Bengtsson (2015), Verma, Khatravath, and Salour (2013) and Verma and Subramanian (2012) all conduct CBA of CBM but include cost associated with lost production time. In all the listed studies, the system being repaired has an associated cost. However, this study focuses on the direct cost of component repairs and system failure.

## 2.3. CBM and Investment Costs

CBM is not triggered by time like TBM or failure like CM. Instead, CBM strategies use condition monitoring technology and perform maintenance when the component's conditions degrade and pass a threshold value prior to failure (Zhao et al 2019; Rausand & Hoyland, 2003). This has been called "just-in time" maintenance (Compare, Antonello, Pinciroli, and Zio, 2022). However, CMS are not perfect and there is often some probability of detection. By performing maintenance based on the component's conditions, maintenance personnel seek to avoid unnecessary maintenance actions (Peng, Dong, and Zuo, 2010), avoid compounded failure effects and associated costs, and reduce downtime (Prajapati et al. 2012). By creating a system that allows maintenance to be performed only when required, the systems can operate with lower maintenance costs and higher overall equipment availability (OEA) (Rastegari, & Bengtsson, 2014). Márquez, Lewis, Tobias, and Roberts (2008) describe CBM benefits as the improvement in reliability, availability, maintainability, and safety (RAMS). However, despite growing research and all the possible advantages, the primary disadvantages are the requirement to develop potentially complex processes for diagnosis and prognosis of failure as well as the associated high initial setup cost (Prajapati et al. 2012; Rastegari, & Bengtsson, 2014; Berdinyazov et al. 2011).

Applying CBM to an engineering system requires both hardware and software units to produce the necessary data to adequately plan maintenance actions. However, achieving a successful CBM system is a complex and expensive endeavor (Rastegari & Bengtsson, 2015). Some of the primary expenses are the stated hardware and software units in addition to training, particular knowledge, and measuring devices (Teixeira et al. 2020; Ahmad & Kamaruddin, 2012). According to Rastegari and Bengtsson (2015) and Teixeira et al. (2020), implementing CBM requires the "selection of the components to be monitored, identification of monitoring techniques and technologies, installation of the required technological means and definition of appropriate data analysis methods." Prior to researching and implementing CBM strategies on a technical system the "organizational, financial, and technical perspective" should be assessed, considered, and used to justify the investment into CBM (Rastegari & Bengtsson, 2014). These complexities and expenses make organizations hesitant to expend the time, money, and resources required to research and implement all the various elements of CBM into their systems. In industry, the vast complexity of CBM has limited its application (Dui, Xu, Zhang, and Wang, 2023).

## 2.4. CBA Modeling

Sandborn and Lucyshyn (2023) state that: "CBA provides a framework to assess the combination of costs and benefits associated with a particular decision or course of action." An ideal CBA performs an extensive analysis of both costs and benefits and considers long-term and indirect effects. Some examples of CBA approaches are CA, ROI, LCC (Sandborn & Lucyshyn, 2023), economic analysis (Berdinyazov et al. 2011), and cost-effectiveness analysis (De Carlo & Arleo, 2013; Rastegari & Bengtsson, 2015). These approaches are methods decision makers can use to evaluate potential options and examine the potential gains.

Sandborn and Lucyshyn (2023) provide an example of combining CBA techniques. Equation 1 from Sandborn and Lucyshyn (2023) shows how CA can be calculated by comparing the difference in the original LCC, $LCC_o$, from the new LCC, $LCC_n$, as shown in Equation 1. The calculated CA can then be added to an ROI calculation that includes the investment costs of implementing the new strategy, as shown in Equation 2 (Sandborn & Lucyshyn, 2023). The greater the CA or the less the investment in CBM, the greater the ROI.

$$CA = LCC_o - LCC_n \qquad (1)$$

$$ROI = \frac{CA}{Investment\ in\ CBM} \qquad (2)$$

A simple example of determining CA can be found by looking at the difference between a single CM event and PM event from De Carlo and Arleo (2013) who provide a description and breakdown of the cost of a single CM and PM event. The cost of both CM and PM according to De Carlo and Arleo (2023) is made of labor costs, material costs, and the costs of lost production. However, the cost of lost production is greater for CM because there is additional time needed to activate maintenance personnel, diagnose the failure, and retrain the system. These three sources of additional time do not occur with a PM event. Therefore, every failure that is preventatively repaired saves money by not paying for unnecessary additional lost production time that comes with failure. This example is the comparison of costs between a single CM and PM event. Systems comprised of multiple, repairable components will have a far more complicated CBA process.

## 2.5. MDFTA and MCS

Fault trees are logic models that are used to represent the system's logic and represent the various combinations and means in which the top event, system failure, can occur (Modarres, Kaminskiy, and Krivtsov, 2016). A fault tree does not model every failure event in a system, only those that contribute to system failure. A fault tree is comprised of events and gates which demonstrate the system's logic. There exists many forms of events and gates. Descriptions of various static and dynamic gates can be found in many places in the literature. By constructing a system's fault tree, both qualitative and quantitative analyses can be performed to determine a variety of aspects about the system such as identifying the system's minimal cut sets and a reliability analysis (Rausand & Hoyland, 2003). A minimal cut set is the smallest set of units whose failure will result in the top event, system failure, occurring (Rausand & Hoyland, 2003).

Traditional fault trees fail to capture the failure behavior of some systems with dynamic behavior when operating. Dynamic fault trees (DFT) capture the dynamic system behavior by adding additional gates. Those additional gates are priority AND (PAND), sequence enforcing (SEQ), standby or spare (SPARE), and functional dependency (FDEP) gates (Gascard & Simeu-Abazi, 2018; Rao et al. 2009). DFTA can be used to evaluate dynamic, multi-component engineering systems for determining aspects about the system reliability such as frequency of system failure, system availability, and unreliability (Gascard & Simeu-Abazi, 2018; Rao et al. 2009).

Gascard and Simeu-Abazi (2018) and Rao et al. (2009) both use MCS to conduct a DFTA. MCS are a valuable tool when solving real world engineering problems, particularly when dealing with complex systems and monetary values of maintenance and operations. MCS can be used to simulate the system's actual processes, something analytical methods struggle with or are incapable of doing. Gascard and Simeu-Abazi (2018) and Rao et al. (2009) use MCS to generate failure times of the DFT basic events and then evaluate if the system, based on the basic event failure times completes a desired operating period or mission time. The process is repeated for many iterations and the number of successful missions is counted and used to assess reliability information about the entire systems. However, both articles analyze the success of a single mission. Their authors don't consider the effect of multiple missions performed in sequential order. Single mission analysis doesn't consider maintenance impacts over a long period of operations. By assessing the system over many missions, the impact of maintenance would become more apparent.

However, fault trees and their analysis are not without challenges. According to Gulati and Dugan (1997): "fault trees lack the modeling power and its solution time increases exponentially with the size of the system being modeled." Aslansefat et al. (2020) provide a summary of both qualitative and quantitative analysis techniques for DFTs. Aslansefat et al. (2020) discuss the various pros and cons associated which the different analysis methods. However, when dealing with the impact of CBM, only a few techniques support repair modeling. Those techniques that support repair modeling are Bayesian Networks (BNs), Generalized stochastic Petri nets (GSPNs), and MCS. Both BNs and GSPN have the potential for state-space explosion. MCS is also a time-consuming operation but improvements such as modulization have helped to improve computational performance (Aslansefat et al. 2020).

A general description of the process of conducting MCS for DFT quantitative analysis is given by Aslansefat et al. (2020). The process involves completing a series of iterations of the system over a desired mission time. During the iterations, failure of components and the system are counted. Once all iterations have been simulated, the system's reliability can be evaluated.

### 2.6. Background Summary

The background presented in this section shows that the use of unmanned systems is rapidly growing while these systems are facing unique maintenance challenges, when conducting CBA for new maintenance approaches, such as CBM. Limited maintenance opportunities are an example of one of these maintenance challenges. However, by considering system-wide impact of maintenance, and degraded operations a CBA can be tailored to provide better analysis of which maintenance strategies will be the most cost effective.

MDFTA and MCS are analysis tools that can be used to model and provide potential solutions to maintenance problems. This study uses these tools due to their abilities to handle systems with multiple components and with unknown impacts from repeated component level maintenances.

## 3. METHODOLOGY

This section will present the aim of the problem to be solved, assumptions made and the proposed approach. The approach contains two portions, one on the CBA and the other on the MDFTA.

### 3.1. Aim and Assumptions

The problem that this study aims to solve is selecting the best CBM strategy for a multi-component, unmanned system. The problem is solved by determining which potential maintenance strategy has the best ROI when compared to the current maintenance strategy.

The assumptions used in this study for solving the problem are as follows:

1. Component failure rates are known.

2. All components are operating, aging, and wearing unless the component has failed, or the entire system has failed. If a component's subsystem fails but the component is working, it will continue to operate even though the subsystem is no longer capable of performing its intended function.

3. Maintenance is only conducted when the system has completed the mission, or the system has failed and been recovered. All maintenance actions only address a single component at a time. No maintenance is conducted during operations (Komianos, 2018; Chang et al. 2021; Dreyer & Oltedal, 2019).

4. PM through CBM is conducted just before the mission where failure would occur (Compare et al. 2022).

5. System failure cost and the cost of degraded operations are independent of the component causing failure or degradation (Yang et al. 2023).

6. Both CM and PM are conducted for a maximum number of minimum repairs. After the maximum number of minimum repairs has been exceeded the component is replaced. This captures the impacts of repeated maintenance actions but does not allow potential failure interarrival times to become so small the system is unable to complete remaining missions.

7. Investment and maintenance costs are assumed based off similar values from other unmanned system maintenance studies (Yang et al. 2023; Dui et al. 2023). All PM costs are less expensive than CM costs and there are no setup costs. Investment costs of CMS are inversely proportional to the components mean time to failure (MTTF) (Kim, An, and Choi, 2017).

8. Supply and storage costs for spare parts are not considered assuming CA is computed for replacing the same components in the original and new life cycle costs.

9. Maintenance costs are constant and do not change over time.

### 3.2. Process: Cost-Benefit Analysis (CBA)

The CBA considered in this study is a combination of CA and ROI (Sandborn & Lucyshyn, 2023). First, we will start with LCC determination. The LCC of system maintenance will be based on the number of maintenance activities $N_{cm}$ and $N_{pm}$ multiplied by the cost per activity $C_{cm}$ and $C_{pm}$. For a multi-component system, different components will have different costs associated with their maintenance activities. This is shown in Equation 3 where subscript $i$ is for the component number and $N_{comp}$ is the total number of components.

Multi-component systems have both critical and non-critical components. Over the operating lifetime of the system, it will have system failures due to critical component failures and failures due to the failure of combinations of non-critical components. If the system does fail, there will be a cost for recovering and restoring the system. The cost of system failure is $C_{f,sys}$ and $N_{f,sys}$ is the number of system failures.

The next piece to consider are the costs associated with lost operational time and time degraded. Lost operational time comes from any difference between the amount of time the system was desired to operate and the time the system operated. Degraded operational time is the portion of time where the system was operating with failed non-critical components or subsystems. Lost operational time costs all the operational value whereas degraded time costs only a portion of the value. The operational time lost is $T_L$ and is the different between the systems desired operating life, $T_{life}$, and the system's actual operational time, $T_{op,sys}$. The degraded time is $T_d$. The cost per unit time of time lost is $C_{OP}$ and the cost per unit time of operating degraded: $d \times C_{OP}$.

The above-listed sources of costs can be used to form the LCC for maintenance of an unmanned system, Equation 3.

$$LCC = (C_{f,sys})(N_{f,sys}) +$$
$$\sum_i^{N_{comp}} \left[ (C_{cm,i})(N_{cm,i}) + (C_{pm,i})(N_{pm,i}) \right] + \qquad (3)$$
$$(C_{OP})(T_L) + (d \times C_{OP})(T_d)$$

However, to get the CA when comparing two strategies, we must determine the difference in LCCs from each maintenance strategy, as provided in Equation 1. The final piece to add is the investment cost. Investment cost is needed to evaluate the ROI of implementing the new maintenance strategy. If a component has a CMS implemented then there will be some investment cost, $C_{inv,i}$. However, if a component does not receive a CMS, then $C_{inv,i}$ is zero. The sum of investment costs is the total CMS investment cost for the system for the new strategy. Adding the investment costs yields Equation 4 for determining ROI of the new strategy compared to the current one. If the addition of a CMS to the system has general costs not associated with a specific component, then those costs would be accounted for in Equation 4's denominator.

$$ROI = \frac{CA}{\sum_i^{N_{comp}} C_{inv,i}} \qquad (4)$$

Equation 4 is the CBA formulation for determining ROI. Based on Equations 1, 3 and 4, the values needed from the Lifetime Simulator (next subsection) for both the old and new maintenance strategies to determine the ROI are: $N_{cm,i}, N_{pm,i}, N_{f,sys}, T_d,$ and $T_{op,sys}$.

## 3.3. Process: Lifetime Simulator

The general layout of the proposed approach, the Lifetime Simulator, is shown in Figure 1. The approach requires two loops. The inner loop, or the Mission Loop, which evaluates the system over its predetermined operating lifetime, $T_{life}$, for several missions. The system is expected to complete several missions over $T_{life}$ with a mission length of $T_m$. The number of missions the system must perform, $N_m$, is determined by $T_{life}$ divided by $T_m$. The number of missions, $N_m$, is the number of iterations of the Missions Loop. The outer loop, or the Iterations Loop, evaluates, after many iterations, the mean and standard deviation of various aspects of the system's maintenance requirements. The number of iterations of the outside loop is $N$. The current iteration number of the Mission Loop and Iteration Loops is designated as $g$ and $h$, respectively.

The next several paragraphs will describe the steps and substeps of the process simulating the system's operational life and estimating the resulting maintenance requirements. The simulation is done using MDFTA with MCS.

### Step 0: Setup and Decompose the DFT to Determine Modules

In this step the entire system DFT will be setup, analyzed and decomposed into modules. A module is a portion of the DFT that can lead to the occurrence of the top event, system failure. Module sizes can range from a single component to several components. A critical component is an example of a single component module. The decomposition analysis in this step seeks to identify the components and subtrees of the entire DFT that will lead to system failure. $M_j$ is the set of components, $i$, contained in module $j$, and $N_{mod}$ is the total number of modules for the system. Every component in the system's DFT will belong to a module.

### Step 1: Initialize Iteration Counters (Time-based): Set $g = h = 0$

### Step 2: Start Iteration Loop (Outer Loop)

The outer loop is to conduct several iterations of the inner loop to provide a better estimation of the system's maintenance requirements. The inner loop conducts MCS to estimate the uncertainty and impact of repeated maintenance events over many repeated missions. The outer loop repeats the process of the inner loop many times to find the mean and standard deviation of the maintenance requirements for the system. To begin the outer loop, the following parameters are initialized by setting them all equal to zero: $T_{op,sys}, T_d, T_{op,i}, t_{f,i}, t_{op,i}, N_{f,sys}, N_{m,complete}, N_{cm,i}, N_{pm,i}, N_{f,j}, N_{r,i},$ and $R_i$.

### Step 3: Generate Initial Failure Interarrival Times: Set $h = h+1$

This step is to generate the first failure interarrival time, $t_{f,i}$, for all components. Generating the failure interarrival times is accomplished with MCS. Each failure interarrival time is generated using a new random number, $U$. $R_i$ is the repair age of the component. Every minimum repair conducted adds to the component's repair age until the component is replaced at which time the component's repair age is set back to zero. Considering the growing repair age for subsequent repairs, the failure time is expected to follow a non-homogenous Poisson process (NHPP).

$$t_{f,i} = F_i^{-1}(1 - U \times [1 - F_i^{-1}(R_i)]) - R_i \qquad (5)$$

Since $R_i$ is zero for all components initially, Equation 5 can be simplified for generating the first failure interarrival times:

$$t_{f,i} = F_i^{-1}(1 - U) \qquad (6)$$

$F_i^{-1}$ is the inverse cumulative distribution function (CDF) and is unique to each component based on their unique distribution parameters, $\alpha_i$ and $\beta_i$.

*Step 4: Start Missions Loop (Inner Loop): Set g = g+1*

The aim of the inner loop is to simulate the lifetime operations of a single system, attempting to complete several missions, $N_m$. The missions are performed one after another, regardless of the prior mission's failure or success. During the inner loop, components will continue to use until they fail, and that is when the maintenance is conducted between the missions. The inner loop captures the impact of repeated operations and component level maintenance for a single system and estimates the resulting maintenance requirements. To being the inner loop, initialize the following parameters by setting them to zero: $t_{m,i}$, $t_{k,j}$, $t_{d,sys}$, and $S_i$.

*Step 5: Assess for CBM*

In Step 5 every component with prior operational time, $t_{op,i} \neq 0$, is evaluated for PM based on the condition monitoring of the component on previous missions and CBM strategy applied. If a component has a CMS, then CMS evaluates the conditions of the component and determines if the component can complete the next mission. If the component is not capable of completing the next mission, then the CMS will alarm, and maintenance personnel will conduct a PM activity for that component. However, the CMS is not perfect and there is a chance that the component will fail the next mission but not be detected by the CMS. The process of assessment for PM through CBM is explained in the following sub-steps:

$$T_m + t_{op,i} < t_{f,i} \qquad (7)$$

*Sub-step 5.1: Assess CMS if component can complete next mission*

$T_m$ is the time of mission and $t_{op,i}$ is the current operational time on component $i$. If $T_m + t_{op,i}$ is less than the failure interarrival time, as shown in Equation 7, then the component can complete the next mission and the CMS does not alert maintenance personnel. If this is the case, no further sub-steps are needed for component $i$. However, if $T_m + t_{op,i}$ is greater than or equal to $t_{f,i}$ then component will fail on the next mission. If this is case move on to sub-step 2.

*Sub-step 5.2: Assess if CMS alarms*

If the component fails the next mission, the next step determines if the CMS detects this. Determining if detection occurs is done by generating a random number, $U$, and comparing it to the CMS detection probability, $P_{CMS,i}$. If the random number is less than $P_{CMS,i}$ then the CMS detected the pending failure, move on to sub-step 3. However, if $U$ is greater than $P_{CMS,i}$, then the CMS did not detect the failure, and PM does not occur.

*Sub-step 5.3: Account for PM*

If the CMS detected failure, then PM was performed. The following values are updated: $T_{op,i} = T_{op,i} + t_{op,i}$, $R_i = R_i + t_{op,i}$, $N_{pm,i} = N_{pm,i} + 1$, and $t_{op,i} = 0$. Accumulated operational time, $T_{op,i}$, and the repair age, $R_i$, are updated based on the components operational time, $t_{op,i}$. The number of PM events on component $i$ is increased by one. Finally, the component's operational time is reset.

*Sub-step 5.4: Generate Next Failure Interarrival Time*

The final sub-step when assessing for PM is to generate a new failure interarrival time for the repair component using Equation 5.

*Step 6: Assess Starting Components*

For Step 6, all starting components attempt to complete the mission. Starting components are all components active at the start of the mission, this excludes components such as cold spares. Like the PM evaluation discussion in Step 5, every staring component is evaluated for completing the mission using Equation 7. For spare components, the dormancy factor, $q_i$, must be considered as shown in Equation 8.

$$q_i \times T_m + t_{op,i} < t_{f,i} \qquad (8)$$

If the adding the mission time, $T_m$, to the components current operational time, $t_{op,i}$, is less than the failure interarrival time, $t_{f,i}$, then the component completes the mission. However, if not then the component will fail the mission and its status is updated to zero, $S_i = 0$. If all starting component statuses are working, $S_i = 1$, continue to Step 7. Otherwise go to Step 8.

*Step 7: Complete Mission*

All starting components are working, $S_i = 1$, then the mission was successfully completed. Equations 9-12 are used to update system and component data. Equation 9 updates the system's accumulated operational time. Equation 10 updates the operational time for starting components. Equation 11 updates the operational time for starting components with dormancy factors greater than 0, like warm spares. Equation 12 accounts for the system completing the mission.

$$T_{op,sys} = T_{op,sys} + T_m \qquad (9)$$

$$t_{op,i} = t_{op,i} + T_m \qquad (10)$$

$$t_{op,i} = t_{op,i} + q_i \times T_m \qquad (11)$$

$$N_{m,complete} = N_{m,complete} + 1. \qquad (12)$$

If $g < N_m$, meaning there are more missions to complete, return to Step 4. If $g = N_m$ go to Step 11.

*Step 8: Assess Modules*

If some of the starting components failed to complete the entire mission, then the consequences to the system must be evaluated. The MDFTA assesses system success based on whether all the modules continued working throughout the mission time. If all modules continued working, then the system continued working. If any modules failed, then the system failed. In this step, any module, $M_j$, with a failed component is assessed to determine if the entire module failed. The assessment for module failure looks at the components of each individual module's subtree. If the subtree is determined to have failed, the module failure time, $t_{k,j}$, is updated with when during the mission the module failed. A module's subtree can contain multiple gates, basic events, and intermediate events.

For static gates, such as AND, OR, and VOTING gates, assessment of whether the gate failed is straightforward. For an OR gate, if any events occurred, the gate failed. For an AND gate, if all events have occurred, then the gate has failed. For a VOTING gate, if the required number of events have occurred, then gate has failed. Figure 1 contains a description of all static gates.

However, for dynamic gates, determining failure is more of a challenge. For example, for a SPARE gate, the gate only has a chance of failing if the primary component, A, has failed. If component A has failed, then the spare component, component B, must complete the remainder of the mission time, or the gate fails. For a FDEP gate, both components must continue working for the gate to work. Failure of component B and the gate fails. Failure of component A and component B can no longer complete its intended task, and the gate fails.

It is possible during this step that multiple modules have been identified to have failed and thus multiple $t_{k,j}$ values have been determined. However, only module one can fail for each mission and thus the system failure time, $t_{f,sys}$, is equal to the minimum of all $t_{k,j}$ that are not zero during this step. If one or modules did fail, and $t_{f,sys}$ has been identified, then the number of module failures is updated using Equation 13. If one or more modules failed, continue to Step 9. If no modules failed, go to Step 10.

$$N_{f,j} = N_{f,j} + 1 \qquad (13)$$

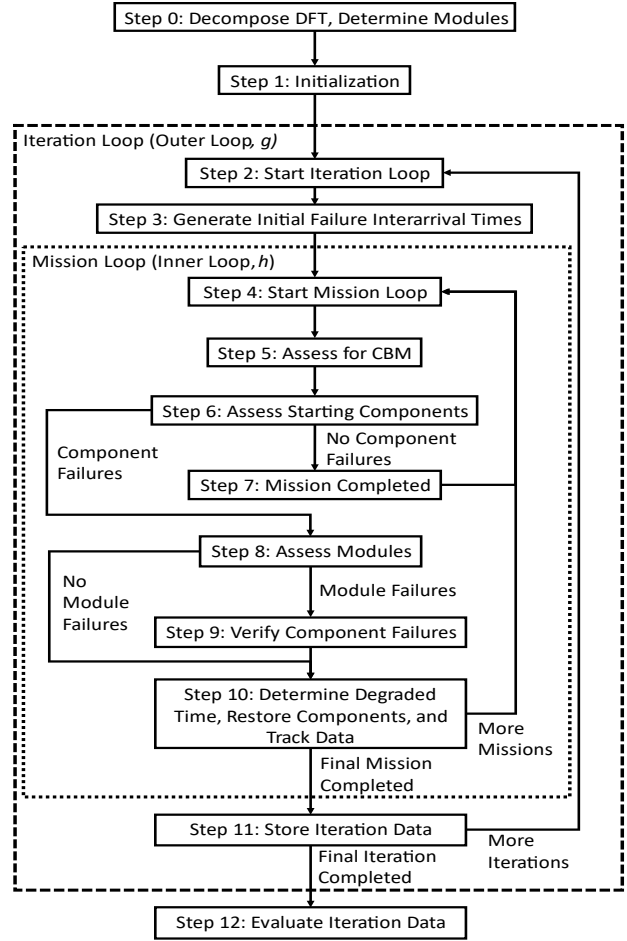

Figure 1: Proposed Approach for the Lifetime Simulator

*Step 9: Verify Component Failures*

If system failure has occurred and $t_{f,sys}$ has been identified, then all failed components, $S_i = 0$, must be checked to verify they truly did fail during this mission. What this step assesses is since the system failed the mission, did component level failures occur before or after the system failure time, $t_{f,sys}$. If the failure occurred before the system failure time, $t_{f,sys}$, then component failed this mission. However, if the component failed after the system failed then component incorrectly failed the mission, and its status must be corrected. The following sub-steps outline this verification of failure process.

*Sub-step 9.1: For all failed components, determine the components time of failure*

Determining the failure time starting and dormant warm components, $t_{m,i}$, is based on the difference between their respective failure interarrival time, $t_{f,i}$, and the current operating time, $t_{op,i}$. Equation 14 shows how starting component mission failure times are determined and Equation 15 shows how the mission failure time of spare components with a dormancy factor that do not become active.

$$t_{m,i} = t_{f,i} - t_{op,i} \qquad (14)$$

$$t_{m,i} = (t_{f,i} - t_{op,i}) \times \frac{1}{q_i} \qquad (15)$$

However, for both warm and cold spare who become active and operate, the time of failure is more complex. The time to failure requires the failure time of the primary component, $t^*_{m,i}$. Equation 16 shows how the mission failure time of active spares occurs after the failure time of the primary component.

$$t_{m,i} = t^*_{m,i} + (t_{f,i} - t_{op,i}) \qquad (16)$$

Once a time of failure has been determined for all failed components, continue to the next sub-step.

*Sub-step 9.2: Evaluate components for failure*

Following sub-step 1, a mission failure time, $t_{m,i}$, has been generated for all failed components. This sub-step evaluates if components failed before or after the system failure time $t_{f,sys}$. If the component failed after the system failure time, $t_{m,i} > t_{f,sys}$, then the component's status is updated back to working, $S_i = 1$, because the components failure did not occur. If the components failure time is equal to less than the system failure time, $t_{m,i} \leq t_{f,sys}$, no action is required.

*Step 10: Determine Degraded Time, Restore Components, and Track System and Component Data*

At this point in the mission simulation, whether system failure has occurred, the time of system failure (if applicable), and which components failed and when failure occurred during the mission have all been evaluated and determined. Hours of degraded operations must now be calculated as well as accounting for repairs and restoration of failed components.

*Sub-step 10.1: Determine system degraded time*

This calculation is done by finding the failure time of the first non-critical component, $t_{m,i}$. The time the first non-critical component failed is equal to $t_{d,sys}$, which is the time on the current mission the system began operating degraded. The time on the current mission the system operated degraded is added to the accumulated time operating degraded. If the mission was completed, the accumulated system degraded time is calculated using Equation 17. However, if system failure occurred, then Equation 18 is used to determine additional degraded time.

$$T_d = T_d + T_m - t_{d,sys} \qquad (17)$$

$$T_d = T_d + t_{f,sys} - t_{d,sys} \qquad (18)$$

*Sub-step 10.2: Restore components and track data*

This is the final step of the mission loop when component failures occurred. In this step, operational time and maintenance data will be updated for the system and all components. Correct accounting of operational time ensures that the impact of repeat missions and prolonged operations are properly accounted for. However, the updates are dependent on whether the system completed the mission and whether the component is working. For the system data, if the mission was completed, system data can be updated using Equations 9 and 12. However, if the system failed use Equations 19 and 20 to account for accumulated operating time and the number of system failures.

$$T_{op,sys} = T_{op,sys} + t_{f,sys} \qquad (19)$$

$$N_{f,sys} = N_{f,sys} + 1 \qquad (20)$$

The following subsections describe how updates are done based on three possible scenarios for the components: (1) Mission completed, working components. (2) System failure, working components. (3) Failed components.

*Mission Completed, Working Components*

The first subsection deals with components that are still working and the system completed the mission. Component data for working, starting components can be updated using Equation 10. However, spare component updates are dependent on whether the spares became active. If a spare component did not become active, Equation 11 can be utilized. However, if a spare component did become active during the mission, then Equation 21 is to be used.

$$t_{op,i} = t_{op,i} + q_i \times t^*_{m,i} + (T_m - t^*_{m,i}) \qquad (21)$$

*System Failure, Working Components*

The next subsection deals with components that are still working but the system failed the mission. Component data for working, starting components can be updated using Equation 22. However, spare component updates are dependent on whether the spares became active. If a spare component did not become active, Equation 23 can be utilized. However, if a spare component did become active during the mission, then Equation 24 is to be used.

$$t_{op,i} = t_{op,i} + t_{f,sys} \qquad (22)$$

$$t_{op,i} = t_{op,i} + q_i \times t_{f,sys} \qquad (23)$$

$$t_{op,i} = t_{op,i} + q_i \times t^*_{m,i} + (t_{f,sys} - t^*_{m,i}) \qquad (24)$$

*Failed Components*

The final subsection covers the updates for failed components. Unlike the previous subsections which only account for the additional operational time of the components, this subsection must also account for the repairs necessary to restore components. For all failed components, the time of failure during the mission, $t_{m,i}$, has already been determined. Therefore, the first step is accounting for the additional accumulated operational time prior to failure. For starting components this is accomplished with Equation 25 and for spares this is done with Equations 26 and 27 with Equation 27 used if the spare became active. The time spent operating prior to failure must also be considered for repair age, $R_i$, as shown in Equations 28-30 for starting components, dormant spares, and active spares, respectively.

$$T_{op,i} = T_{op,i} + t_{op,i} + t_{m,i} \tag{25}$$

$$T_{op,i} = T_{op,i} + t_{op,i} + q_i \times t_{m,i} \tag{26}$$

$$T_{op,i} = T_{op,i} + t_{op,i} + q_i \times t_{m,i}^* + (t_{m,i} - t_{m,i}^*) \tag{27}$$

$$R_i = R_i + t_{op,i} + t_{m,i} \tag{28}$$

$$R_i = R_i + t_{op,i} + q_i \times t_{m,i} \tag{29}$$

$$R_i = R_i + t_{op,i} + q_i \times t_{m,i}^* + (t_{m,i} - t_{m,i}^*) \tag{30}$$

Once the operational time has been added for accumulated operational time and repair ages, all failed components undergo repairs. The number of repair events and minimum repairs are accounted for using Equations 31 and 32. The operational time is set back to zero, $t_{op,i} = 0$ following repairs.

$$N_{cm,i} = N_{cm,i} + 1 \tag{31}$$

$$N_{r,i} = N_{r,i} + 1 \tag{32}$$

If $N_{r,i} > N_{r,max,i}$ then the repair age is reset back zero. After $N_{r,max,i}$ minimum repairs the component is replaced, vice repaired. This resets the repair age as a new component is installed. This process of conducting several minimum repairs and an eventual replacement allows the methodology to capture the impacts of repeat minimum repairs but does not allow the next failure interarrival time to become so small that mission completion becomes unlikely. The final step is generating a new failure interarrival time, $t_{f,i}$. This is done using Equation 5.

If $g < N_m$ return to Step 4. If $g = N_m$ continue to Step 11.

*Step 11: Storing Iteration Data*

To account for any operational time on component still functioning on the final mission, the current operational time must be added to the accumulated time, as shown in Equation 33.

$$T_{op,i} = T_{op,i} + t_{op,i} \tag{33}$$

After executing the final mission, a single iteration of the Iteration Loop has been completed. The data from the Mission Loop is stored in a database for future use and assessment. The data stored in the data base is the following variables:

$T_{op,sys,h}, T_{op,i,h}, T_{d,h}, T_{d,h}, N_{f,sys,h}, N_{cm,i,h}, N_{pm,i,h}, N_{m,c,h},$ and $N_{f,j,h}$.

If $h < N$ return to Step 2. If $h = N$ continue to Step 12.

*Step 12: Evaluating Iteration Data*

Once the Iteration Loop has been completed, the database of stored information is evaluated. The evaluation is done by determining mean and standard deviation of the values stored in the database as shown in Figure 1. A sample mean and standard deviation determination is shown with Equations 34 and 35.

$$\bar{T}_{op,sys} = \frac{\sum_h^N T_{op,sys,h}}{N} \tag{34}$$

$$S(T_{op,sys}) = \left[ \frac{\sum_{h=1}^N (T_{op,sys,h} - \bar{T}_{op,sys})^2}{N} \right]^{1/2} \tag{35}$$

## 4. Case Study and Results

This section will demonstrate the application of the proposed methodology to a USV case study from the literature (Gao et al. 2021).

### 4.1. Case Study Setup

The USV is comprised of 5 subsystems and 71 total components. The five subsystems are: power, communication, navigation, navigation control, and environment data acquisition. The USV's DFTs contain OR, AND, FDEP, SPARE, and VOTING gates. The DFTs also contain cold spare (CSP) and warm spare (WSP) components. The USV DFTs are shown in Figures 2 to 7. Figure 2 shows how the 5 subsystems connect to the overall USV system while Figures 3-7 detail the DFTs of each of the subsystems. These figures are adapted from Gao et al. (2021) and maintain the same ordering of component numbers. Additionally, the acronyms in the subsystem specific DFTs shown in Figures 3-7 can be found in the article by Gao et al. (2021). The module breakdown is from Step 0 of the methodology discussed in Section 3 and is summarized in Table 1. The following paragraphs will detail all the information needed for the methodology.

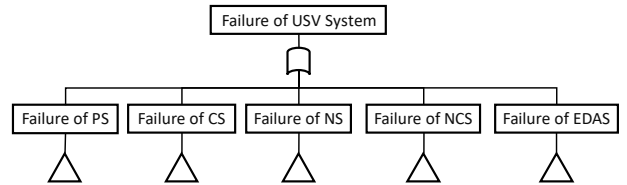

Figure 2: System DFT (adapted from Gao et al. (2021))

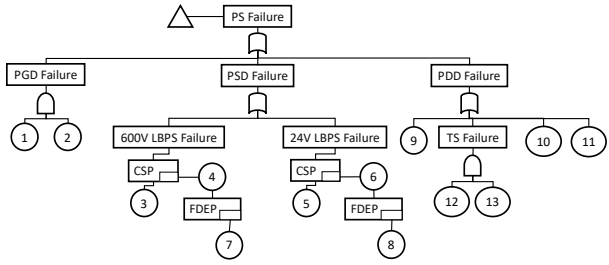

Figure 3: Power Subsystem (PS) DFT (adapted from Gao et al. (2021))

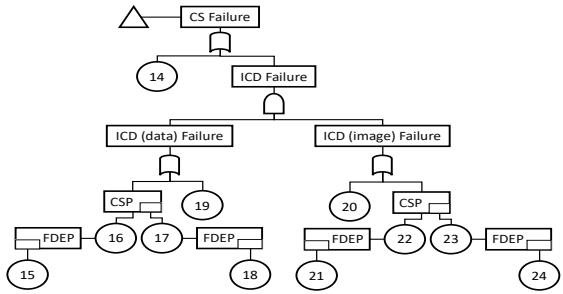

Figure 4: Communication Subsystem (CS) DFT (adapted from Gao et al. (2021))

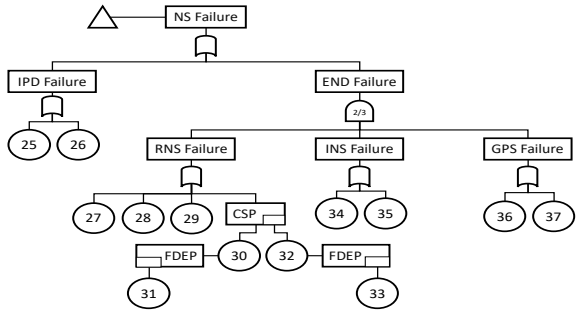

Figure 5: Navigation Subsystem (NS) DFT (adapted from Gao et al. (2021))

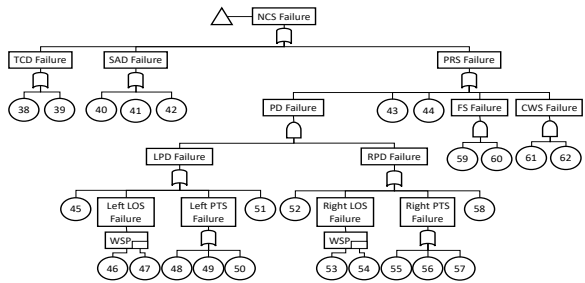

Figure 6: Navigation Control Subsystem (NCS) DFT (adapted from Gao et al. (2021))

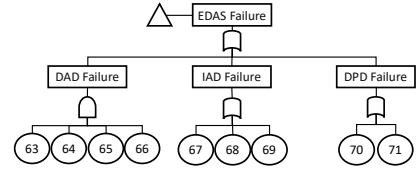

Figure 7: Environment Data Acquisition Subsystem (EDAS) DFT (adapted from Gao et al. (2021))

| Module # | Comp. # | Module # | Comp. # |
|---|---|---|---|
| 1 | 9 | 15 | 40 |
| 2 | 10 | 16 | 41 |
| 3 | 11 | 17 | 42 |
| 4 | 1,2 | 18 | 43 |
| 5 | 12,13 | 19 | 44 |
| 6 | 3, 4, 7 | 20 | 59, 60 |
| 7 | 5, 6, 8 | 21 | 61, 62 |
| 8 | 14 | 22 | 45-58 |
| 9 | 15-24 | 23 | 67 |
| 10 | 25 | 24 | 68 |
| 11 | 26 | 25 | 69 |
| 12 | 27-37 | 26 | 70 |
| 13 | 38 | 27 | 71 |
| 14 | 39 | 28 | 63-66 |

Table 1: USV Module Breakdown

System operating life, mission lengths, and iteration data are: $T_{life} = 200{,}000\ hours$, $T_m = 200\ hours$, and $N = 1000$. The number of missions, $N_m$, is 1000. The dormancy factor, $q_i$, for cold spares is zero and 0.5 for warm spares. Failure data is contained in Tables 2 and 3. The failure data is from EPRD2014 and NPRD2016. However, the failure data from EPRD2014 and NPRD2016 is for an exponential distribution. To utilize a NHPP and model impacts of minimum repairs, the failure data was converted for use with a Weibull distribution with a shape parameter slightly greater than one. A shape parameter slightly greater than one will keep the model close to an exponential distribution ($\beta = 1$) but add the element of an increasing failure rate with time and repairs. Therefore, all components have the same shape parameter, $\beta = 1.2$, and a scale parameter that gives the same MTTF as the exponential failure data. Minimum repairs will be conducted for the first 5 repairs, then the component will be replaced.

For investment and maintenance costs, no specific values are given in the study by Gao et al. (2021) However, other studies by Yang et al. (2023) and Dui et al. (2023) for unmanned system provide cost data for components in comparable systems. Using the values for CM and PM, values are approximated for this study and given in Tables 2 and 3. System failure costs and costs per hour of operation are also not given. The study by Yang et al. (2023) considers the costs for recovery of an AMS following failure and thus provides comparable values for this study of a USV. For hourly operations costs, unlike a production system, the system is not generating profit during its operations. Instead, the value of each hour is based on the total cost of the system and its operating life. O'Rourke (2019) provides information on the United States Navy's USV program. In the study by O'Rourke (2019), the value of a medium sized USV is listed at $35,000,000. Using this value and the operating life of 200,000 hours, an hour of operations can be valued at $175. System costs are summarized in Table 4. Finally, since this study is the first of its kind, there was no specific data given for investment cost for research, development, and installment of a component's CMS. For this study, the investment cost of a CMS is assumed but is inversely correlated with the components shape parameter. The smaller the shape parameter, the smaller the MTTF, the greater the costs of a CMS.

| Component Number | Failure Rate ($hrs^{-1}$) | Scale Parameter ($hrs^{-1}$) | CM Cost ($) | PM Cost ($) | CM Investment Cost ($) |
|---|---|---|---|---|---|
| 1 | $1.150 \times 10^{-5}$ | 92,440 | 12,000 | 8,000 | 15,300 |
| 2 | $1.150 \times 10^{-5}$ | 92,440 | 12,000 | 8,000 | 15,300 |
| 3 | $3.773 \times 10^{-3}$ | 280 | 1,000 | 500 | 100,000 |
| 4 | $3.773 \times 10^{-3}$ | 280 | 1000 | 500 | 100,000 |
| 5 | $1.680 \times 10^{-4}$ | 6330 | 800 | 600 | 75,500 |
| 6 | $1.680 \times 10^{-4}$ | 6330 | 800 | 600 | 75,500 |
| 7 | $8.707 \times 10^{-7}$ | 1,220,960 | 1,500 | 500 | 13,700 |
| 8 | $8.707 \times 10^{-7}$ | 1,220,960 | 1,500 | 500 | 13,700 |
| 9 | $4.515 \times 10^{-6}$ | 235,460 | 4,000 | 1,000 | 14,300 |
| 10 | $4.515 \times 10^{-6}$ | 235,460 | 4,000 | 1,000 | 14,300 |
| 11 | $2.131 \times 10^{-5}$ | 49,890 | 2,500 | 1,200 | 16,900 |
| 12 | $4.672 \times 10^{-5}$ | 22,750 | 5,000 | 3,000 | 21,900 |
| 13 | $4.672 \times 10^{-5}$ | 22,750 | 5,000 | 3,000 | 21,900 |
| 14 | $8.456 \times 10^{-7}$ | 1,257,200 | 10,000 | 2,000 | 13,700 |
| 15 | $5.629 \times 10^{-7}$ | 1,888,590 | 7,500 | 5,000 | 13,700 |
| 16 | $5.793 \times 10^{-5}$ | 18,350 | 3,000 | 1,800 | 24,600 |
| 17 | $5.793 \times 10^{-5}$ | 18,350 | 3,000 | 1,800 | 24,600 |
| 18 | $5.629 \times 10^{-7}$ | 1,888,590 | 7,500 | 5,000 | 13,700 |
| 19 | $3.482 \times 10^{-7}$ | 3,053,100 | 8,000 | 7,000 | 13,700 |
| 20 | $3.482 \times 10^{-7}$ | 3,053,100 | 8,000 | 7,000 | 13,700 |
| 21 | $5.629 \times 10^{-7}$ | 1,888,590 | 7,500 | 5,000 | 13,700 |
| 22 | $5.793 \times 10^{-5}$ | 18,350 | 3,000 | 1,800 | 24,600 |
| 23 | $5.793 \times 10^{-5}$ | 18,350 | 3,000 | 1,800 | 24,600 |
| 24 | $5.629 \times 10^{-7}$ | 1,888,590 | 7,500 | 5,000 | 13,700 |
| 25 | $2.606 \times 10^{-5}$ | 40,790 | 2,200 | 2,000 | 17,800 |
| 26 | $5.497 \times 10^{-6}$ | 193,390 | 2,000 | 1,000 | 14,400 |
| 27 | $1.872 \times 10^{-5}$ | 56,790 | 1,000 | 800 | 16,500 |
| 28 | $1.250 \times 10^{-4}$ | 8,500 | 9,000 | 6,000 | 48,700 |
| 29 | $1.407 \times 10^{-4}$ | 7,560 | 5,000 | 2,000 | 57,200 |
| 30 | $1.888 \times 10^{-6}$ | 563,080 | 8,500 | 6,500 | 13,900 |
| 31 | $5.629 \times 10^{-7}$ | 1,888,590 | 7,500 | 5,000 | 13,700 |
| 32 | $1.888 \times 10^{-6}$ | 563,080 | 8,500 | 6,500 | 13,900 |
| 33 | $5.629 \times 10^{-7}$ | 1,888,590 | 7,500 | 5,000 | 13,700 |
| 34 | $4.672 \times 10^{-5}$ | 22,750 | 2,500 | 1,000 | 21,900 |
| 35 | $1.243 \times 10^{-5}$ | 85,530 | 7,500 | 3,000 | 15,400 |

Table 2: Data for Components 1-35

## 4.2. Proposed CBM Strategy #1

The first potential CBM strategy will be to monitor all critical components. The critical components are: 9, 10, 11, 14, 25, 26, 38-44, and 67-71 to a $P_{CMS,i}$ value of 0.5. This CBM strategy will be evaluated against the current CM only strategy. The results are shown in Figure 8. Only components with greater than 10 failures are shown. Figure 8 shows the results of comparing the two strategies.

Figure 8 contains the mean value of the number CM and PM activities performed for each component for each strategy and the associated 95% confidence interval error bars. Using the complete results and mean values of maintenance requirements, the results of Strategy #1 indicate that adding CMS to critical components reduces maintenance costs by $806,200 for an investment cost of $420,700. Additionally, the system operated for 5,610 more hours but 3,022 more hours degraded. The costs of maintenance, lost operational time, and degrade compared to the investment costs results in an ROI of Strategy #1 of 4.53.

| Component Number | Failure Rate $(hrs^{-1})$ | Scale Parameter $(hrs^{-1})$ | CM Cost ($) | PM Cost ($) | CM Investment Cost ($) |
|---|---|---|---|---|---|
| 36 | $1.171 \times 10^{-5}$ | 90,780 | 1,200 | 800 | 15,300 |
| 37 | $1.452 \times 10^{-6}$ | 732,150 | 8,000 | 7,000 | 13,800 |
| 38 | $3.120 \times 10^{-6}$ | 342,710 | 2,000 | 1,800 | 14,000 |
| 39 | $4.753 \times 10^{-6}$ | 223,670 | 2,000 | 1,800 | 14,300 |
| 40 | $6.239 \times 10^{-6}$ | 170,390 | 800 | 150 | 14,500 |
| 41 | $1.248 \times 10^{-5}$ | 85,180 | 800 | 150 | 15,500 |
| 42 | $1.095 \times 10^{-5}$ | 97,090 | 1,000 | 600 | 15,200 |
| 43 | $1.738 \times 10^{-4}$ | 6,120 | 1,500 | 800 | 80,100 |
| 44 | $1.738 \times 10^{-4}$ | 6,120 | 1,500 | 800 | 80,100 |
| 45 | $1.537 \times 10^{-4}$ | 6,920 | 12,000 | 8,000 | 65,300 |
| 46 | $1.961 \times 10^{-5}$ | 54,210 | 6,000 | 2,000 | 16,600 |
| 47 | $1.961 \times 10^{-5}$ | 54,210 | 6,000 | 2,000 | 16,600 |
| 48 | $1.086 \times 10^{-6}$ | 978,900 | 8,000 | 3,500 | 13,800 |
| 49 | $1.268 \times 10^{-6}$ | 838,400 | 4,500 | 4,000 | 13,800 |
| 50 | $1.248 \times 10^{-5}$ | 85,180 | 4,500 | 4,000 | 15,500 |
| 51 | $8.933 \times 10^{-4}$ | 1,190 | 6,000 | 2,000 | 90,000 |
| 52 | $1.537 \times 10^{-4}$ | 6,920 | 12,000 | 8,000 | 65,300 |
| 53 | $1.961 \times 10^{-5}$ | 54,210 | 6,000 | 2,000 | 16,600 |
| 54 | $1.961 \times 10^{-5}$ | 54,210 | 6,000 | 2,000 | 16,600 |
| 55 | $1.086 \times 10^{-6}$ | 978,900 | 8,000 | 3,500 | 13,800 |
| 56 | $1.268 \times 10^{-6}$ | 838,400 | 4,500 | 4,000 | 13,800 |
| 57 | $1.248 \times 10^{-5}$ | 85,180 | 4,500 | 4,000 | 15,500 |
| 58 | $8.933 \times 10^{-4}$ | 1,190 | 6,000 | 2,000 | 90,000 |
| 59 | $1.429 \times 10^{-4}$ | 7,440 | 6,000 | 2,000 | 58,500 |
| 60 | $1.429 \times 10^{-4}$ | 7,440 | 6,000 | 2,000 | 58,500 |
| 61 | $1.545 \times 10^{-4}$ | 6,880 | 6,000 | 2,000 | 65,800 |
| 62 | $1.545 \times 10^{-4}$ | 6,880 | 6,000 | 2,000 | 65,800 |
| 63 | $1.166 \times 10^{-5}$ | 91,170 | 5,000 | 4,500 | 15,300 |
| 64 | $1.243 \times 10^{-5}$ | 85,530 | 5,000 | 4,500 | 15,400 |
| 65 | $6.075 \times 10^{-5}$ | 17,500 | 4,000 | 2,500 | 25,300 |
| 66 | $3.801 \times 10^{-5}$ | 27,970 | 3,000 | 2,200 | 20,100 |
| 67 | $1.854 \times 10^{-5}$ | 57,340 | 1,500 | 800 | 16,400 |
| 68 | $1.854 \times 10^{-5}$ | 57,340 | 1,500 | 800 | 16,400 |
| 69 | $8.456 \times 10^{-7}$ | 1,257,200 | 10,000 | 2,000 | 13,700 |
| 70 | $3.740 \times 10^{-6}$ | 284,250 | 2,200 | 2,000 | 14,100 |
| 71 | $9.268 \times 10^{-5}$ | 11,470 | 3,500 | 1,500 | 35,000 |

Table 3: Data for Components 36-71

However, looking at Figure 8, it can be observed that some critical components such as 9, 10, 14, 26, 38-42, 69, and 70 had an insignificant number of failures compared to other non-critical components such as 3, 4, 51, and 58. Therefore, the next strategy will remove the CMS from the listed critical components with minimal failures and add a CMS to the listed non-critical components.

### 4.3. Proposed CBM Strategy #2

The second strategy adjusts the components being monitored. As previously stated, the results from Strategy #1 highlight that some components did not need a CMS while others did. The results of the new strategy compared to the current one of CM only are shown in Figure 9.

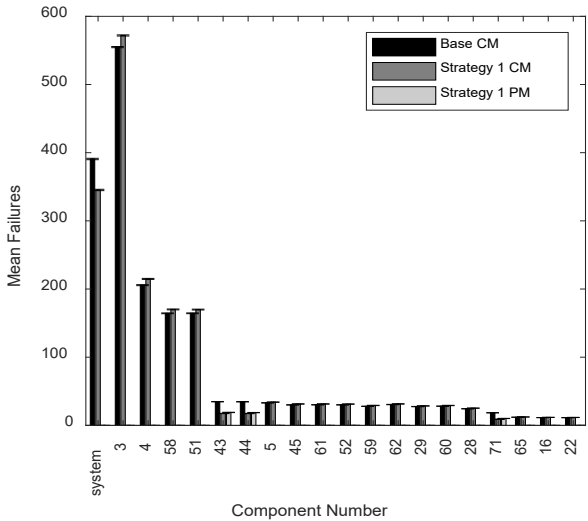

Figure 8: Bar graph with 95% confidence interval error bars comparing the CM and PM activities between the base strategy and Strategy #1

Once again, using the mean values of maintenance requirements, the results of Strategy #2 show that adding CMS to components 3, 4, 11, 25, 43, 44, 51, 58, 67, 68, 70, and 71 reduces maintenance costs by $3,151,500 for an investment cost of $642,700. Additionally, the system operated for 16,490 more hours but for 2,001 more hours degraded. The costs of maintenance, lost operational time, and degraded time compared to the investment costs results in an ROI of Strategy #2 of 9.72.

## 5. DISCUSSION

The case study demonstrates the effectiveness of the CBA for CBM decisions. The CBA showed that Strategy #2 would be preferred based on a better ROI. Even though less components were monitored, the components selected for Strategy #2 had a higher investment cost. Yet, the investment was worth it by having a twice as good ROI. However, there are other factors besides the final ROI worth noting.

The first item worth discussing is the connection between components based on maintenance strategies. When comparing the current strategy to Strategy # 1, there is an increase in the number of CM events for some components, such as components 3 and 4. This is because, as CBM was implemented the system operated longer and thus experienced more component level failures in some components. These additional failures would be counterproductive to efforts to avoid unnecessary maintenance costs and thus their impact cannot be ignored. The impact of component level maintenance is captured by

MCS and the MDFTA. These tools allowed us to simulate the impacts of maintenance and capture the changes across components and the system.

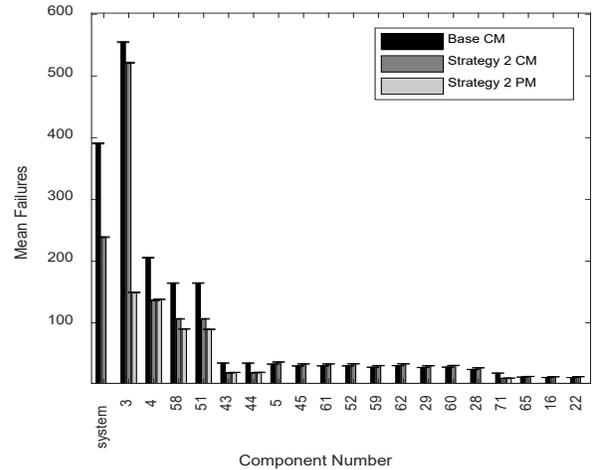

Figure 9: Bar graph with 95% confidence interval error bars comparing the CM and PM activities between the base strategy and Strategy #2

The next item worth mentioning is noticing how CBM impacts some components differently. Components 43 and 44 had a 0.5 $P_{cms}$ value under both strategies. In both strategies, the number of component failures for components 43 and 44 were reduced by about 50%. However, observing component 3 we can observe the change in failures due to the addition of a 0.5 $P_{CMS}$ CMS was not a 50% reduction. This is likely due to component 3 having very small MTTFs compared to other components in the system. Therefore, even if maintenance is conducted on component 3, there is a considerable probability it may fail the next trip regardless of recent repairs.

| System lifetime | $T_{life}$ | 200,000 hours |
|---|---|---|
| Mission time | $T_m$ | 200 hours |
| Number of missions | $N_m$ | 1,000 |
| Number of iterations | $N$ | 1,000 |
| Cost of system failure | $C_{sys,f}$ | $20,000 |
| Cost of an operational hour | $C_{OP}$ | $175 |
| Degraded factor | $d$ | 0.2 |

Table 4: System and Mission Data

The next item to discuss is how the final ROI values could be impacted by other topics mentioned in maintenance literature, such as storage or holding costs and discounting. The storage of spare parts can have a cost associated with it and this cost, if included, could change the impact of a new maintenance strategy. The storage costs could be potentially added to the cost avoidance calculations as shown in Equations 3 and 4. For this study, storage costs were assumed to be negligible. However, if storage costs had been included in this study, they likely would have made the ROI larger as the proposed maintenance strategies resulted in the system experiencing less failures and thus less spare parts would have been needed and stored. The result would have been maintenance costs, including the cost of storage being smaller for the new strategy and thus a larger ROI.

Another maintenance impact could be discounting. Discounting is where conducting maintenance on one component is done in tandem with maintenance on a different component since combining the activities would be cheaper than performing each item separately. Discounting was not included in this study. However, including discounting if a system does receive a discount when combining maintenance items would obviously change the results and make them more accurate for that system. Consider this example for combining discounting and CBM and maximizing the impact. Say that components A and B in a system undergo maintenance together, hence are subject to discounting. By using the methodology discussed in this paper, the organization could attempt to add a CMS to only component A or only component B or to both and compare the results. There is a good chance that if the components work together, monitoring one may be sufficient and cost effective compared to monitoring both or none. Discounting and storage costs could be added to the methodology as discussed and improve the results if the organization deems them necessary.

Finally, for verification of the results of the case study, the article by Gao et al. (2021) is used. In the article, the probability of failure from each of the 5 subsystems that comprise the USV are provided. The results are that if system failure occurred, the probability of each subsystem causing failure are 0.612, 0.372, 0.090, 0.024, and 0.0005 for the power, navigation control, environment data acquisition, navigation, and communication subsystems, respectively. In our results, using the same failure rates at Gao et al. (2021) and a replacement only strategy, we obtain the following subsystem failure probabilities: 0.595, 0.319, 0.069, 0.0164, and 0.0005. These values correlate with Gao et al. (2021) and allow us to verify the results generated by the MDFTA for later CBA calculations.

## 6. CONCLUSION

In this paper for the first time a CBA is conducted for adding CMS to an unmanned system while considering unmanned system maintenance challenges, the impact CMSs at the component level has on the entire system, and the cost of degraded system operations. The study combines MDFTA and MCS to estimate the maintenance requirements for an unmanned system, both with and without CBM. The differences in maintenance requirements due to CBM and the associated investment costs are then used to determine the ROI of the new CBM strategy. This study shows how the value of the ROI of various proposed CBM strategies can be determined and used to identify the better strategy. While conducting the CBA to determine the ROI of a new strategy, the approach considered challenges of unmanned systems, accounted for degraded operations, and took a system-wide approach to maintenance impacts.

The presented methodology was applied to a USV case study and determined the ROI for two different CBM strategies when compared to a CM only policy. The results showed that due to limited maintenance opportunities and high costs of system failure, monitoring less components and non-critical components led to a greater ROI. The results of the approach were verified against the previously completed failure results for the USV (Gao et al. 2021).

This study can be implemented by organizations operating unmanned systems and contemplating implementing CMS and CBM to their systems The study gives organizations an analysis tool for evaluating the possible value of CBM, which is of great need in industry, according to Berdinyazov et al. (2011). However, CMS and PHM techniques can also extend beyond maintenance and improve operations of unmanned systems as well. In a study by Hazra, Chatterjee, Southgate, Weiner, Growth, and Azarm (2024), the authors present a novel framework for optimizing operational profiles of a USV using PHM techniques. This shows how CMS and PHM can have value outside of just maintenance.

This study is unique by incorporating unmanned system challenges like high system failure costs and no maintenance personnel present during operations in system simulations and overall CBA. Additionally, the presented study is very flexible and allows for extensive sensitivity analysis. The flexibility would enable users to include variations in cost data, maintenance strategies, and cost grouping factors which would improve the accuracy of the study and make it more practical.

Future works on this topic can include exploration into many areas covered by the study. Optimization is one area that could be further explored. Optimization could occur for the CBA if there was some way to demonstrate the cost of different levels of performance of the CMS for different components. In other words, if the investment costs for CMS not only depended on the component but the detection accuracy as well, then optimization could identify which components to monitor and to what degree of accuracy to achieve the greatest return.

### ACKNOWLEDGEMENT

This study was supported in part by the Office of Naval Research (ONR) under grant number N000142212459. This support does not constitute an endorsement by the funding agency of the opinions expressed in the paper.

### ACRONYMS

| | |
|---|---|
| BN | Bayesian network |
| CA | Cost avoidance |
| CBA | Cost-benefit analysis |
| CBM | Condition-based maintenance |
| CDF | Cumulative distribution function |
| CEA | Cost-effectiveness analysis |
| CM | Corrective maintenance |
| CMS | Condition monitoring system |
| DFT | Dynamic fault tree |
| DFTA | Dynamic fault tree analysis |
| GSPN | Generalized stochastic Petri nets |
| LCC | Life cycle cost |
| MASS | Maritime autonomous surface ships/systems |
| MCS | Monte Carlo simulation |
| MDFTA | Modular dynamic fault tree analysis |
| MTTF | Mean time to failure |
| NHPP | Non-homogeneous Poisson process |
| PHM | Prognostics and health management |
| PM | Preventative maintenance |
| ROI | Return on investment |
| SHM | Structural health monitoring |
| TBM | Time-based maintenance |
| UAV | Unmanned aerial vehicle |
| USV | Unmanned surface vessel |
| UV | Unmanned vehicle |

**NOMENCLATURE**

$C_{cm,i}$ Cost of a corrective maintenance event for component $i$ ($)

$C_{pm,i}$ Cost of a preventative maintenance event for component $i$ ($)

$C_{inv,i}$ Investment cost for a condition monitoring system for component $i$ ($)

$C_{OP}$ Cost per hour of operations ($)

$C_{f,sys}$ Cost of system failure ($)

$d$ Degraded factor, portion of lost hourly value due to degraded operations

$g$ Missions loop iteration counter

$h$ Iterations loop iteration counter

$i$ Subscript for component number

$j$ Subscript for module number

$M_j$ Sets of components $i$ contained in module $j$ (Array)

$N_{comp}$ Number of components

$N_{mod}$ Number of modules

$N_{f,sys}$ Number of system failures

$N_{f,sys,h}$ Number of system failures for iteration $h$

$N_{f,j}$ Number of failures for module $j$ (Array)

$N_{f,j,h}$ Number of failures for module $j$ for iteration $h$

$N_m$ Number of missions for the system to attempt

$N_{m,c}$ Number of missions completed successfully

$N_{m,c,h}$ Number of missions completed successfully for iteration $h$

$N_{cm,i}$ Number of CM for component $i$ (equivalent to the number of failures) (Array)

$N_{cm,i,h}$ Number of CM for component $i$ for iteration $h$

$N_{pm,i}$ Number of PM for component $i$ (Array)

$N_{pm,i,h}$ Number of PM for component $i$ for iteration $h$

$N_{r,i}$ Number of minimum repairs conducted on component $i$

$N_{r,max,i}$ Maximum number of minimum repairs for component $i$

$P_{CMS,i}$ Probability of detection for the CMS of component $i$

$q_i$ Dormancy factor of component $i$

$R_i$ Repair age of component $i$

$S_i$ Status of component $i$ (1=working, 0=failed) (Array)

$t_{d,sys}$ Time on the current mission the system begins operating degraded

$t_{f,i}$ Failure interarrival time of component $i$ (Array)

$t_{f,sys}$ Time on the current mission the system failed

$t_{k,j}$ Time on the current mission failure of module $j$ occurs (Array)

$t_{m,i}$ Time on the current mission component $i$ fails (Array)

$t_{m,i}^*$ Failure time of a primary component in a spare gate

$t_{op,i}$ Accumulated operating time of the current component $i$ (Array)

$T_d$ Accumulated time the system spent degraded

$T_{d,h}$ Accumulated time the system spent degraded for iteration $h$

$T_{life}$ Desired operating lifetime of the system

$T_L$ Lost operational time for the system

$T_m$ Length of individual missions

$T_{op,i}$ Accumulated operating time of component, $i$ (Array)

$T_{op,i,h}$ Accumulated operating time of component $i$ for iteration $h$

$T_{op,sys}$ Accumulated operating time of system

$T_{op,sys,h}$ Accumulated operating time of the system for iteration $h$

$U$ Random number generated for MCS and CMS [0,1]

$\alpha_i$ Scale parameter of component $i$

$\beta_i$ Shape parameter of component $i$